# DNA, Human Memory, and the Storage Technology of the 21st Century

Masud Mansuripur

Optical Sciences Center, The University of Arizona, Tucson, Arizona 85721

**Abstract.** The sophisticated tools and techniques employed by Nature for purposeful storage of information stand in stark contrast to the primitive and relatively inefficient means used by man. We describe some impressive features of biological data storage, and speculate on approaches to research and development that could benefit the storage industry in the coming decades.

**Introduction.** The storage of information is ubiquitous in our technological society: paper, film, semiconductor memories, audio/video-tapes, magnetic/optical disks, etc., collectively contain many petabytes of information. In contrast, Nature has been frugal in its use of information storage techniques. Blueprints of life, both of plant and of animal, are stored in the DNA molecules.[1-3] Pre-programmed (i.e., instinctive) as well as learned information reside in the nervous systems of higher animals.[4,5] The human immune system stores information about past pathogens in the form of primed lymphocytes (e.g., T-cells and B-cells), using this information to mass-produce and rapidly deploy antibodies and specialized immune cells when an old pathogen reappears.[6,7] These instances aside, it is hard to find purposeful employment of data storage in Nature.

Despite their rarity, the natural mechanisms of information storage are extremely powerful and versatile. A complement of chromosomes not only contains the entire description of a plant or an animal, but it also carries the step-by-step instructions for building the individual from a single initial cell. The human brain can store vast amounts of information embodied in images, sounds, scents, event sequences, and abstract concepts to which an individual may be exposed through a lifetime. The brain forms automatic links among the stored data, recalls by association, and responds to external events by exploiting its reservoir of pre-programmed and learned data-bases.

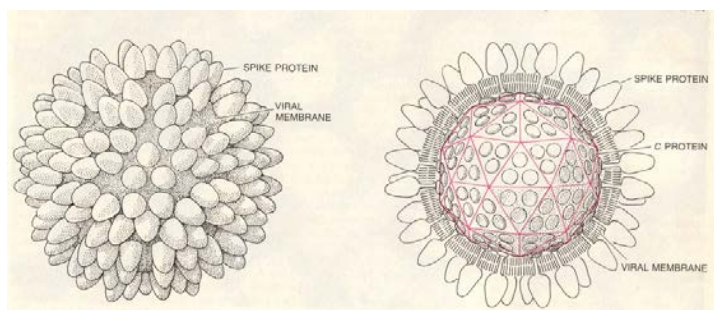

**Figure 1.** The SFV particle is 65 nm in diameter. The RNA inside the capsid is a chain of 12,700 nucleotides, the order of which provides the information needed for making the viral proteins. The capsid consists of 180 molecules of a protein, forming a regular polygon with 60 triangular faces. The outer viral membrane is composed of two layers of lipid molecules, with their hydrophobic heads facing outward and their hydrophilic tails facing inward. Inserted into this membrane are 180 spikes, each made up of three linked proteins.[8]

To cite just one example of efficient data storage by Nature, consider the Semliki Forest Virus (SFV) depicted in Fig. 1. The virus consists of a 12,700 nucleotide-long strand of RNA, wrapped in a coat of exquisitely arranged protein molecules and surrounded by a spike-protein studded lipid membrane. The entire virus is 65 nm in diameter, smaller than the tiniest pit on a Digital Versatile Disk (DVD), yet the information content of its RNA molecule is 25.4 kbits (2 bits per nucleotide). When this virus enters a host cell, it sheds its membrane and protein coat, and takes over the machinery of the host cell to make copies of its own RNA and proteins. The host cell eventually assembles the mass-produced components of the SFV, and sends them out at the rate of several thousand per hour to infect other cells within the host organism.[8]

This paper reviews some of the fundamental aspects of cellular processes and neuronal behavior in biological systems, with an eye toward the mechanisms responsible for information storage and retrieval. Recent years have witnessed a steady convergence of digital electronic storage and computation or information processing. It should not come as a surprise, therefore, to find convergence, even seamless merger, of storage and processing in biological systems as well. Admittedly, we are far from the ultimate goal of employing biomaterials and our knowledge of biological processes in the service of building better information storage devices. It is our modest intention, however, to point out the power and flexibility of biological information processing, and to encourage the reader to explore this path in search of alternative modes and means for future technologies. For the reader unfamiliar with the language of the biologist, a glossary of relevant terms and concepts is included at the end of the paper.

**Information Stored in Chromosomes**. A typical bacterial chromosome consists of a double-helix ring of deoxyribonucleic acid (DNA), making up tens or hundreds of genes;[9] see Fig. 2. (Bacteria are prokaryotes, that is, cells without nuclei. The bacterial chromosome floats in the cell's cytoplasm, surrounded by the cell membrane and a protective shell.) All the proteins and enzymes needed by the bacterium are encoded by these genes, and are made by the protein-fabricating machinery of the cell.[10-12] The DNA molecule, therefore, contains the blueprint for the construction of the entire organism. The basic operating principle is the same in fungi, plants, and animals – all of which belong to the class of eukaryotes and consist of nucleated cells;[9] see Fig. 3. Eukaryotes, however, have more chromosomes (for example, 4 pairs in fruit flies, 23 pairs in human beings), and each chromosome is longer and may contain more genes (on average, a few thousand genes per human chromosome).[10]

In sexually-reproducing organisms there are pairs of each chromosome (i.e., the homologous chromosome pair), each member of the pair containing the same gene – or different alleles of the same gene – at identical loci. Of the two members of a given pair of chromosomes, one is contributed by the mother, the other by the father. Different alleles of a given gene (residing at a specific locus along a given chromosome) encode for the same protein or enzyme, but the alleles are biochemically different, and could therefore give rise to structurally and functionally different proteins/enzymes. In some cases only one of the two alleles (either maternal or paternal) is expressed, while the other copy of the gene remains dormant (recessive). In other cases both genes (residing, as they are, at identical loci of a homologous pair) are expressed. In a sexually-reproducing organism, each of the two copies of a given gene has a 50% chance of being transmitted to the offspring in a gamete (i.e., sperm or egg cell).[1,10]

At any given time during the life of a cell, some of the genes are copied onto messenger ribonucleic acid or mRNA (see Fig. 4) and transported to the cytoplasm, outside the cell nucleus.[10] There they line up the amino-acids (the same 20 amino-acids are used in all living organisms) with help from ribosomes, transfer RNA, and specific enzymes, and create the proteins encoded in the genes. At the same time, some of the products of the genes (enzymes, proteins) return to the nucleus, and prevent other genes from expression, say, by blocking their pairing with mRNA molecules; see Fig. 5. The information stored in the genes, therefore, not only encodes for specific proteins, but it also enables some of these proteins to return to the nucleus to act on other genes, thereby modifying their behavior. The chemistry of the environment also plays a significant role in determining which genes are expressed in a given cell during a given phase of development of the organism.[13,14]

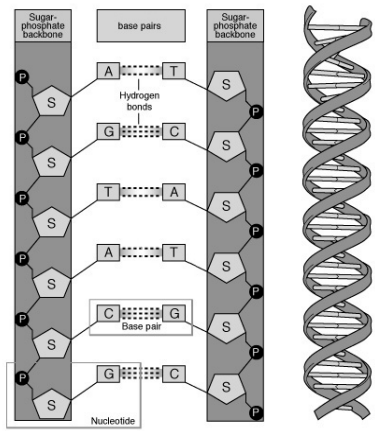

**Figure 2. Deoxyribonucleic acid (DNA)**. Built from nucleic acid bases, sugars, and phosphates, the double-stranded molecule is twisted into a helix. Each spiraling strand, comprised of a sugar-phosphate backbone and attached bases, is connected to a complementary strand by non-covalent hydrogen bonding between paired bases. The bases are adenine (A), thymine (T), cytosine (C) and guanine (G). A and T are connected by two, and G and C by three hydrogen bonds. The right-handed double helix has ~10 nucleotide pairs per helical turn. The double-helix structure of DNA was discovered in 1953 by James Watson and Francis Crick. (Reproduced with permission from: http://www.nhgri.nih.gov/.)

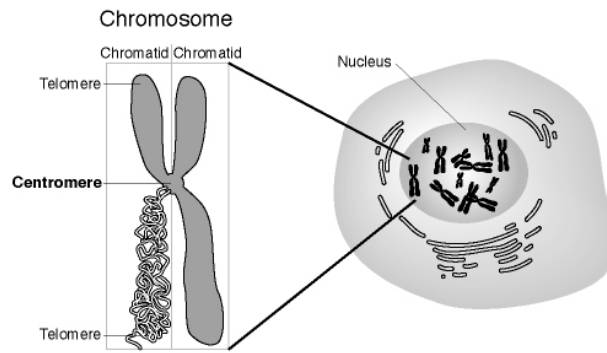

**Figure 3**. In eukaryotes the chromosomes reside in the nucleus of the cell. In sexually-reproducing organisms every chromosome comes in a pair, one member of the pair from the mother, the other from the father. During normal cell division (mitosis), each chromosome is copied, and the daughter cells receive a copy of each and every chromosome, thus receiving a full complement of all chromosomal pairs. During germ cell (gamete) production (meiosis) the homologous chromosome pairs within the nucleus come together and recombine (i.e., shuffle their genes), before the cell splits to form individual gametes. The gametes receive only half of the total number of chromosomes, one from each pair.[10] (Reproduced with permission from: http://www.nhgri.nih.gov/.)

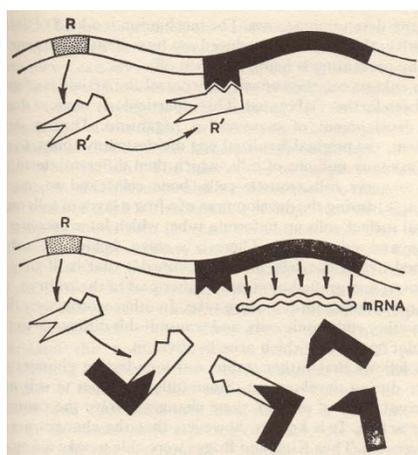

**Figure 5**. Gene-activating mechanism. A gene, shown in black, is prevented from acting by a repressor substance R' produced by the gene R. Below, 'inducing' molecules, also shown in black, have entered the cell from outside, and by combining with the repressor substance prevent the repressor from switching the gene off. The gene is therefore producing mRNA.[10]

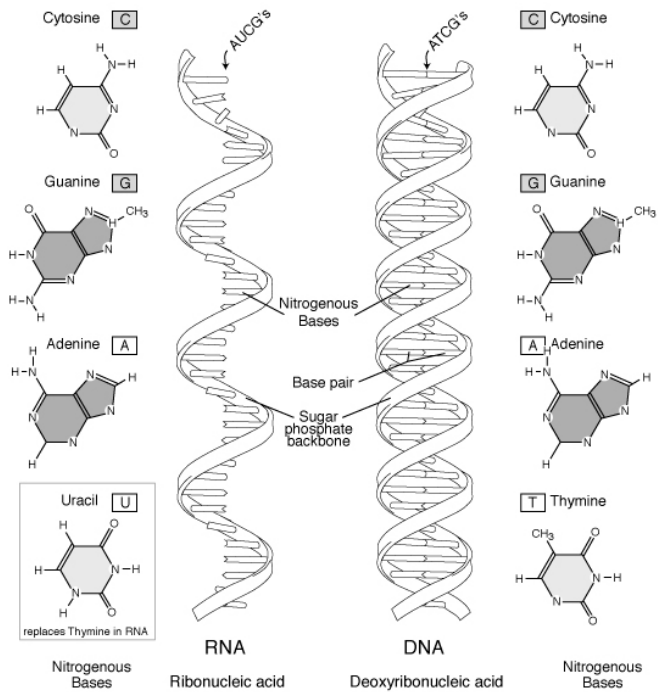

**Figure 4. Ribonucleic acid (RNA)**. A chemical similar to a single strand of DNA. In RNA the letter U (for uracil) is substituted for T in the genetic code. Messenger RNA delivers DNA's genetic message to the cytoplasm of a cell where proteins are made. Three-letter sequences of messenger RNA, known as codons, code for specific amino acids.
(Reproduced with permission from: http://www.nhgri.nih.gov/.)

The bottom line is that the information stored in chromosomes, like a giant software package, has many subroutines that can be invoked under various circumstances, and the next subroutine to be invoked is determined by the results of previous "computations" in a particular environment. (The data fed to the computer program plays the role of the electrochemical environment within which a cell performs its functions). Like a jumbo-jet factory, the body of a complex organism consists of many divisions and subdivisions. A complete copy of the instruction manual for building the jumbo-jet may be present in each and every subdivision, but workers at each location use only the instructions in one or perhaps a few sections of the manual for constructing the part(s) assigned to their subdivision.

**Natural versus Man-made Data Storage**. There are similarities but also significant differences between man-made and natural memories. For example, DNA is a one-dimensional molecule that uses a quaternary base of nucleic acids (Adenine, Cytosine, Guanine, and Thymine) to store a sequence of information along its length. Similarly, information on a CD is recorded along a spiral track using a sequence of 9 symbols (pit lengths = 3T–11T). In both cases special symbols are employed to mark the beginning and the end of a single block of information (a gene, a sector). Error correction coding is employed to ensure the integrity of the information-content against random errors. In a CD various sectors are linked through a file allocation table (FAT) that specifies the logical sequence of these sectors, even though they may not be stored sequentially along the track. Similarly, links exist among various genes, not only on the same chromosome but also on different chromosomes, that specify the sequence of their activation, or prohibit the expression of one gene based on the expression or non-expression of others.

Both the DNA molecule and the spiral track on a CD are one-dimensional although, in principle, there are no reasons why DNA should not be two-dimensional (i.e., sheets containing A, C, G, T or similar molecules connected to their neighbors on all sides). In our 3D world, a 2D sheet can easily replicate itself by acting as the template for constructing a complementary sheet, then separating from its copy, much the same as a single strand of DNA replicates itself. 3D replication based on the same principle, however, is impossible, because of the need for the molecules to interpenetrate, link, and then separate. Nature has chosen 1D strands of DNA over 2D sheets, presumably because the sheets are either difficult to make or because they provide no additional advantages.

The differences between DNA and CD data storage are staggering. The chromosomes are copied in their entirety during each cell division, so that a new cell will have a complete set of chromosomes of the entire individual (except for the gametes, i.e., the egg and sperm cells, which carry only one copy of each chromosome, and also the red blood cells, which have no chromosomes at all). A given cell obviously does not need all the genes, because only a few genes are expressed in each cell; the rest remain dormant. Most of the genes in a given cell are never used; those that are needed are not used all the time, but are typically expressed at certain stages of development of the organism. In contrast, when a file from a CD is copied to another medium, only those sectors that constitute the desired file are transferred; there is no need to copy an entire CD if all one needs is a specific piece of information. Each chromosome consists of a complementary pair of DNA strands, the double helix; although the useful information is only recorded on one of the two strands, both are needed to ensure the success of the reproduction process during cell division. In contrast, data along a CD track is single-stranded, because the copying mechanism does not rely on the existence of a complementary strand.

In sexually-reproducing organisms, the chromosomes come in pairs, with the same gene appearing on two homologous chromosomes (one maternal, the other paternal). Both genes are capable of expression, although in certain cases the end product of one gene may be dominant over that of its counterpart. When one of the genes is dysfunctional (either by inheritance, or because of copying errors during cell division, or because of mutation) the other "allele" of the same gene is usually used by the cell's protein-building machinery, thus avoiding malfunction. A serious problem usually arises when both genes are dysfunctional (e.g., a protein is not produced, or a modified version of it is produced). In contrast, the stored information on a CD typically comes in a single copy, and there are no (functionally similar but physically different) duplicates that could be used if and when the necessity arises.

**Example of Existing Technology: The DNA Chip** (adapted from [15]). By adapting methods of microprocessor manufacturing, Affymetrix, a California-based company, has created microchips that contain thousands of distinct DNA probes on a glass substrate; see Fig. 6. The glass is coated with a grid of tiny spots, ~20μm in diameter, each spot containing millions of copies of a short sequence of DNA. A computer keeps track of the location of each DNA sequence.

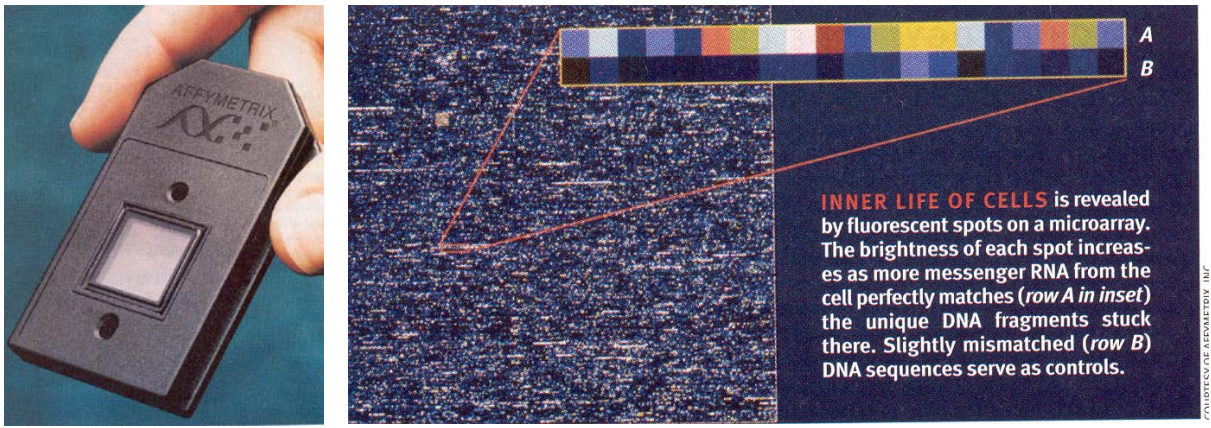

**Figure 6**. DNA chip can sense the state of up to 400,000 genes in a tissue sample. The number of probes on a single glass wafer may soon exceed 60 million, in which case the entire human genome will fit on 200-300 wafers.[15]

To make snapshots of gene activity, the workers extract mRNA from sample cells. Using enzymes, they make millions of copies of these mRNA molecules, tag them with fluorescent dyes, and break them up into short fragments. The tagged fragments are spread over the chip and, overnight, perform a remarkable feat of pattern matching, randomly bumping into the DNA probes fixed to the chip until they stick to one that contains a perfect genetic match. Although there are occasional mismatches, the millions of probes in each spot ensure that it lights up only if complementary mRNA is present. The brighter the spot fluoresces when scanned by a laser, the more mRNA of the kind exists in the cell.

Affymetrix currently makes over 100,000 chips a year, using photosensitive chemicals to build DNA probes one nucleotide at a time. Agilent, Hitachi, and Protogene Laboratories, among others, use modified ink jet printers whose heads squirt A, T, G, and C nucleotides instead of cyan, magenta, yellow, and black inks. Canon is reportedly working with bubble jets to deposit DNA sequences, whereas Corning, Motorola, and Incyte Genomics employ precision robots that place microdroplets of presynthesized sequences onto prepared slides.

The Gene Chip technology clearly demonstrates the feasibility of recording and readout of information using DNA molecules and biochemical principles; the speed and efficiency of the process, however, leave something to be desired.

**Biochemical and DNA-based Nanocomputers** (adapted from [16]). A chemical computer is one that processes information by making and breaking chemical bonds, storing logic states in the resulting molecular structures. A chemical nanocomputer would perform such operations selectively among molecules taken a few at a time, in volumes only a few cubic nanometers. Proponents of biochemical nanocomputers point to an "existence proof" in the commonplace activities of humans and other animals with multicellular nervous systems. Presently, the possibility of artificial fabrication of biochemical computers seems remote because the operating mechanisms of animal nervous systems are poorly understood. In the absence of a deeper understanding, research has proceeded in alternative directions. One line of investigation has sought to adapt naturally occurring biochemicals for use in computing processes that do not have parallels in Nature. Another approach has been to culture and employ living tissues for computational purposes.

In 1994 Leonard Adleman, then at MIT, took a giant step towards fulfilling the promise of artificial biochemical computers.[17,18] He used fragments of DNA to compute the solution to a complex graph theory problem (a version of the traveling salesman problem). As shown in Fig. 7, Adleman utilized DNA base-sequences to represent vertices of a network or graph. Combinations of these sequences, formed randomly by the massively parallel biochemical reactions in test tubes, described random paths through the graph. Using the tools of biochemistry, Adleman extracted the correct answer to the posed problem out of the many random paths represented by the product DNA strands. Like a multiprocessor computer, this type of DNA computer is able to consider many solutions to a problem simultaneously. Moreover, the approximately $10^{23}$ DNA strands employed in such a calculation are orders-of-magnitude greater in number (and more densely packed) than the processors in today's massively parallel electronic computers.

It seemed at first that Adleman's method would be limited to the solution of combinatoric problems. Recent work by R. Lipton at Princeton University has shown, however, that the approach may be applied to a much wider class of digital computations. The problems of fast and efficient input/output, as well as numerous other obstacles, must be overcome before this promising new approach can be broadly applied.[19,20]

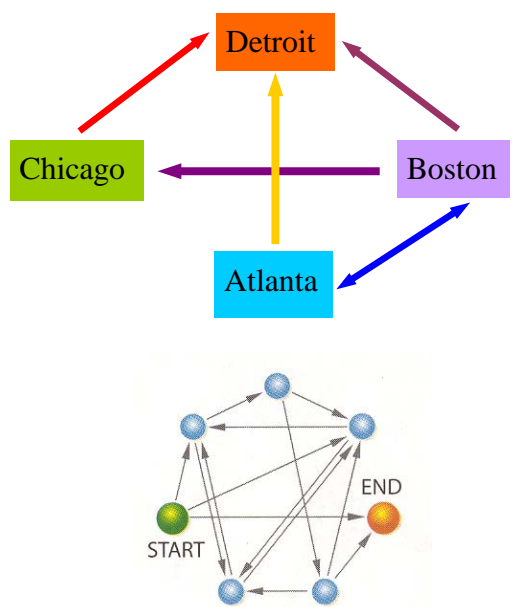

| CITY | DNA NAME | COMPLEMENT |
|---|---|---|
| Atlanta | ACTTGCAG | TGAACGTC |
| Boston | TCGGACTG | AGCCTGAC |
| Chicago | GGCTATGT | CCGATACA |
| Detroit | CCGAGCAA | GGCTCGTT |
| **FLIGHT** | **DNA FLIGHT NUMBER** | |
| Atlanta-Boston | GCAGTCGG | |
| Atlanta-Detroit | GCAGCCGA | |
| Boston-Chicago | ACTGGGCT | |
| Boston-Detroit | ACTGCCGA | |
| Boston-Atlanta | ACTGACTT | |
| Chicago-Detroit | ATGTCCGA | |

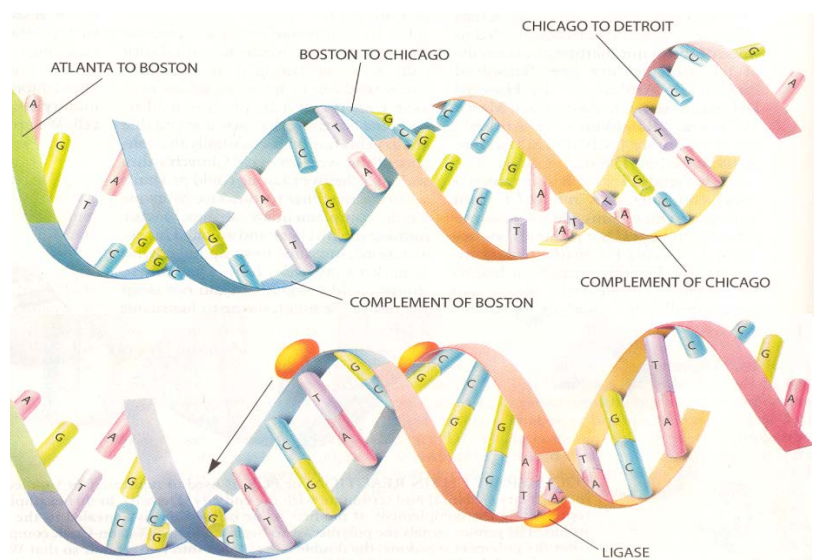

**Figure 7** (Adapted from L. M. Adleman, "Computing with DNA," *Scientific American* **279**, August 1998). The Hamiltonian path problem is exemplified by a map of cities connected by certain nonstop flights. In this example, for instance, one can fly directly from Boston to Chicago but not vice versa. The goal is to determine whether a path exists from the start city of Atlanta to the finish city of Detroit, passing through each of the remaining cities exactly once. In Adleman's scheme each city is assigned a DNA sequence (TCGGACTG for Boston) consisting of a first name (TCGG) and a last name (ACTG). Also shown in the Table are the Watson-Crick complementary city names in which every C is replaced by a G, every A by a T, etc. DNA flight numbers are assigned by concatenating the last name of the city of origin with the first name of the city of destination. For this particular problem only one Hamiltonian path exists, Atlanta → Boston → Chicago → Detroit, represented by the 24-base-long sequence GCAGTCGGACTGGGCTATGTCCGA. (Also shown in this figure is the diagram of seven cities and 14 connecting flights used in the original Adleman's experiment.) Watson-Crick annealing, in which Cs pair with Gs and As join with Ts, results in DNA flight-number strands being held end-to-end by strands encoding the complementary DNA city names. Ligases connect the splinted molecules; wherever the protein finds two strands of DNA in proximity, it will covalently bond them into a single strand.

**Bacteriorhodopsin-based Memories.** Robert Birge of Syracuse University has suggested to use the light-sensitive protein dye bacteriorhodopsin, produced by some bacteria, for memory and logic device applications (see Figs. 8, 9). According to Birge and his collaborators, bacteriorhodopsin can provide a high-density optical memory that can be integrated into an electronic computer to yield a hybrid device of superior performance compared to conventional, purely electronic computers.[21-23]

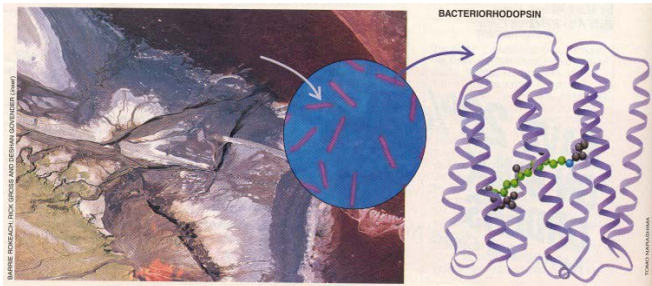

**Figure 8**. Salty waters of Owens Lake, California, shown in an aerial view, have a purple hue caused by the presence of bacteria (inset) containing the colorful protein bacteriorhodopsin. The protein, depicted here as a ribbon, includes a light-absorbing chromophore segment (shown as balls and sticks). When the chromophore is excited by light, its structure changes and thereby alters the conformation of the rest of the protein. Because bacteriorhodopsin adopts different, readily detectable states in response to light, it can serve as logic gates, or switches, in protein-based optical computers.[22]

**Figure 9**. Sequence of structural changes induced by light allows for the storage of data in bacteriorhodopsin molecules. Green light transforms the initial resting state, bR, to the intermediate K, which then relaxes and forms M followed by O. If the O intermediate is exposed to red light, a so-called branching reaction occurs. Structure O converts to P, which quickly relaxes to the Q state, a form that remains stable almost indefinitely. Blue light, however, converts Q back to bR. Any two long-lasting states can be assigned the binary value 0 or 1, making it possible to store information as a series of bacteriorhodopsin molecules in one state or the other.[22]

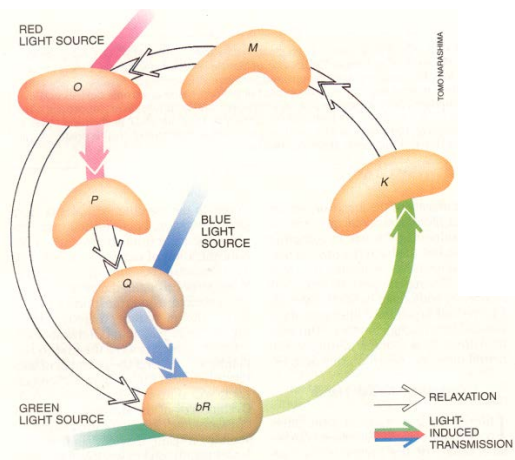

**Polymerase Chain Reaction**. The simple and elegant method of polymerase chain reaction (PCR) for amplifying small amounts of DNA fragments was discovered in 1983 by K. B. Mullis, then at Cetus Corporation.[24] The fragments may come from a human tissue, they may be isolated from a sample of hair or blood recovered from a crime scene, from a woolly mammoth that has been frozen in a glacier for forty thousand years, or from an insect trapped within 80 million year-old amber (fossilized pine resin). The PCR amplification of a single fragment of DNA into billions of identical pieces can be accomplished in just a few hours in a test tube that contains at least one piece of the original DNA (the template), mixed with some simple chemicals and reagents. Figure 10 shows a section of DNA molecule from a chromosome residing inside the nucleus of a cell. By chemical convention, one end of a given strand of DNA is denoted 3' and the other end 5'. Since the two complementary strands are antiparallel, the 3' end of one strand is always paired with the 5' end of the other strand.

One must add a sufficient number of "primer" molecules to the PCR test tube to satisfy the growing needs of the rapidly multiplying DNA throughout the process. A primer molecule (also known as an oligonucleotide probe) is typically made by the techniques of organic chemistry, and contains 20-30 nucleotide bases (A, T, G, C). Each primer must be complementary to an a priori known segment on one strand of the original DNA molecule (the template), which enables the primer to bind to that segment under proper conditions. A separate primer must be furnished in the test tube for each strand of the original template; in other words, two different types of primer are needed for a given template.

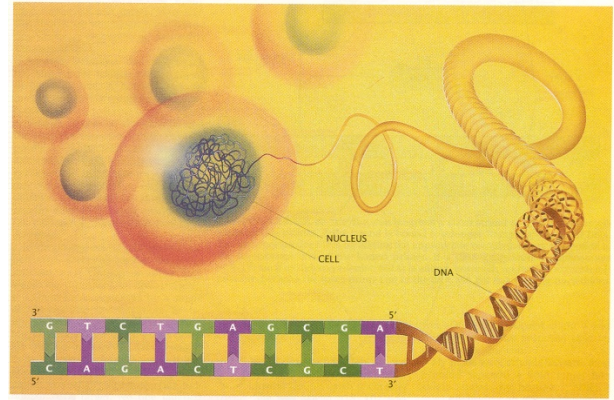

**Figure 10**. DNA consists of two strands of linked nucleotides, the sequence in one strand being complementary to that in the other (i.e., A's are always opposite T's, and G's are opposite C's). Complementarity binds the strands together. Each strand has a 3' and a 5' end. Because their orientations oppose one another, the strands are said to be antiparallel.[24]

Each primer must be designed to bind to a location near the 3' end of its corresponding DNA strand, as shown in Fig. 11. A primer molecule, when attached to a single strand of template DNA, begins to grow at its own 3' end in accordance with the complementary base sequence specified by the attached template. (Of course the solution must contain plenty of A, T, C, G nucleic acids to satisfy the needs of the growing molecules. In addition, there must be present in the solution an abundant supply of DNA polymerase, a natural enzyme that facilitates the duplication of DNA strands.) After a few minutes in the test tube, each primer grows all the way to the 5' end of its associated strand, thus completing the duplication of the original template.

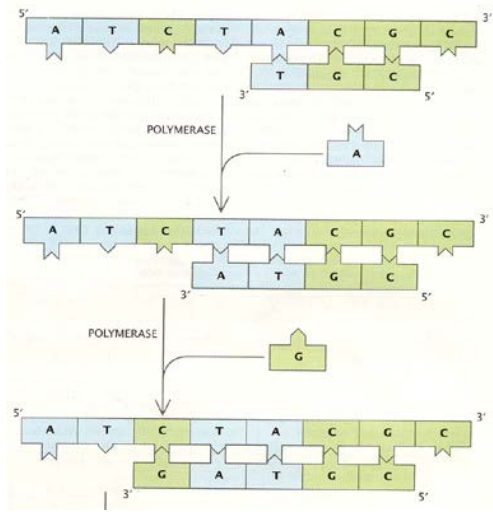

The first step in a polymerase chain reaction is to heat up the test tube containing nucleic acid molecules, an appropriate DNA polymerase, two types of primer for the intended DNA target, and, of course, the original fragment of DNA. (The essential criterion for any DNA sample is that it contain at least one intact DNA strand encompassing the region to be amplified, and that any impurities are sufficiently diluted so as not to inhibit the polymerization step of the PCR.) When heated to ~95°C for 30 seconds, the two strands of the template separate from each other; the DNA molecule is said to be denatured at this point. The temperature is subsequently lowered to ~55°C, where two primer molecules bind (or "anneal") to specific locations on the two (separated) strands from the template; this binding may take about 20 seconds. Once the primers are bound, the temperature of the solution is raised to ~75°C, at which point the DNA polymerase goes to work, adding bases (complementary to those on the corresponding DNA strands) at the 3' end of each primer molecule. The copying of the two strands will be completed in approximately one minute, and the original DNA molecule (the template) will be reproduced in duplicate. The process can be repeated by cycling the temperature, thus doubling the number of DNA molecules at the end of each cycle. The total number of double-stranded DNA molecules in the test tube thus grows exponentially, reaching a billion in just 30 steps.

**Figure 11**. The enzyme DNA polymerase can lengthen a short strand of DNA, called an oligonucleotide primer, if the strand is bound to a longer "template" strand of DNA. The polymerase does this by adding the appropriate complementary nucleotide to the 3' end of the bound primer.[24]

If the DNA polymerase is obtained from aquatic bacteria residing in thermal vents or hot springs (bacteria that can withstand high temperatures), there will be no need to replenish the supply of polymerase at the end of each heating cycle, because it will not degrade by heating. The DNA polymerase extracted from *Thermus aquaticus*, the sultry bacterium from the Yellowstone National Park, is now universally favored for PCR.

DNA polymerases, whether from humans, bacteria, or viruses, cannot copy a chain of DNA without a short sequence of nucleotides to "prime" the process, or get it started. (Cells have an enzyme called a "primase" that actually makes the first few nucleotides of the copy.) This stretch of DNA is called a primer. Once the primer is made, the polymerase can take over making the rest of the new chain.

Primers are annealed to the denatured DNA template to provide an initiation site for the elongation of the new DNA molecule. Primers can be specific to a particular DNA sequence or they can be "universal." The latter are complementary to nucleotide sequences that are common in a particular set of DNA molecules. Thus, universal primers bind to a wide variety of DNA templates. Bacterial ribosomal DNA genes, for example, contain nucleotide sequences common to all bacteria. Thus, universal primers for bacterial DNA can be made by creating primers which are complementary to these sequences. Examples of universal primers for bacteria are:

        Forward 5' GAT CCT GGC TCA GGA TGA AC 3' (20 mer)
        Reverse 5' GGA CTA CCA GGG TAT CTA ATC 3' (21 mer)

Animal cell-lines contain a particular sequence known as the ALU gene. (Approximately 900,000 copies of the ALU gene are distributed throughout the human genome, and multiple copies of the same can be found in the genome of other animals.) The ALU gene thus provides a universal primer for animal cell-lines; this primer is especially useful because it binds in both forward and reverse directions.

    ALU universal primer: 5' GTG GAT CAC CTG AGG TCA GGA GTT TC 3' (26mer)

When using universal primers, the annealing temperature should be lowered to 40-55°C.

**Memory and the Nervous System**. There are an estimated $10^{10}$–$10^{11}$ neurons in the central nervous system of a human being. Although there are many types of neurons of differing sizes and shapes, the principles of operation of these neurons are remarkably similar.[4,25] Figure 12 shows the structure of a typical neuron. The neuron is a specialized cell, consisting of a cell body (soma), where the nucleus resides, a relatively long cable, axon, along which the output signal travels to reach other neurons (or muscle cells, or glands), and many short branches, dendrites, which are in contact with the axonal terminals of other neurons, acting as receiving terminals for their own neuron;[25] see Fig. 13. A typical cell may have hundreds or thousands of dendrites, each no more than a millimeter long. The axon, on the other hand, could be quite long, reaching in length to several feet in some cases. Each axon branches off and makes contact with dendrites of several other neurons.[5,25] The contact between the end-point of an axonal branch and another cell's dendrite or cell body is called a synapse (see Fig. 14). Electrical signals travel down an axon in the form of spikes (pulses) of ~100 mV magnitude and ~1 ms duration, reaching the dendrites unattenuated, as shown in Fig. 15. The travelling speed of these signals is typically of the order of several meters per second (in some cases reaching 100 m/s or even faster).[26]

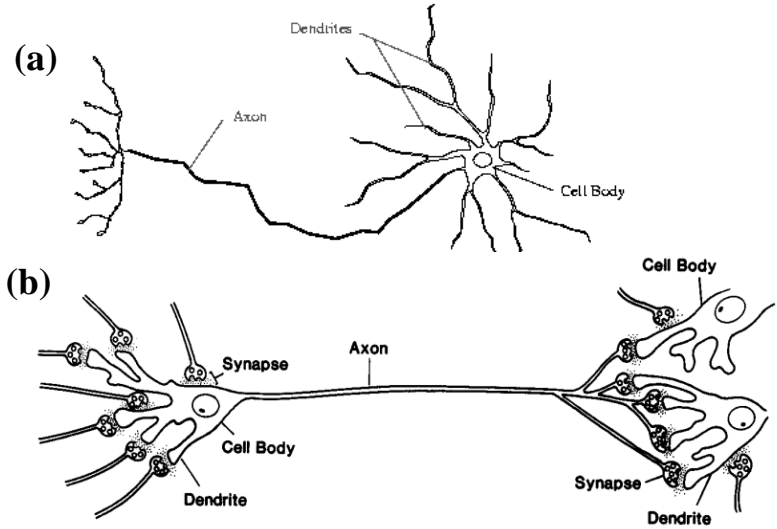

**Figure 12**. (a) The structural unit of the nervous system, a neuron contains a nucleus and cell body from which has grown a tree of many branches and twigs spreading in various directions. A neuron typically has many dendrites and one axon. The dendrites branch and terminate in the vicinity of the cell body. In contrast, axons can extend to distant targets, more than a meter away in some instances. Dendrites are rarely more than about a millimeter long and often much shorter. (From: Synapse Web, Boston University, http://synapses.bu.edu/)
(b) The electrical signals flow into the dendrites and out through the axon. Thus in this diagram the information flows from left to right.

At the synapse, the arriving electrical pulse causes the release of certain chemicals (neuro-transmitters) into the region between the two terminals (the synaptic cleft). These chemicals are picked up by receptors on the postsynaptic neuron and cause the transmission of an electrical signal from the dendrite to the postsynaptic cell. Communication between the connected cells may therefore be characterized as electrical $\rightarrow$ chemical $\rightarrow$ electrical.[5]

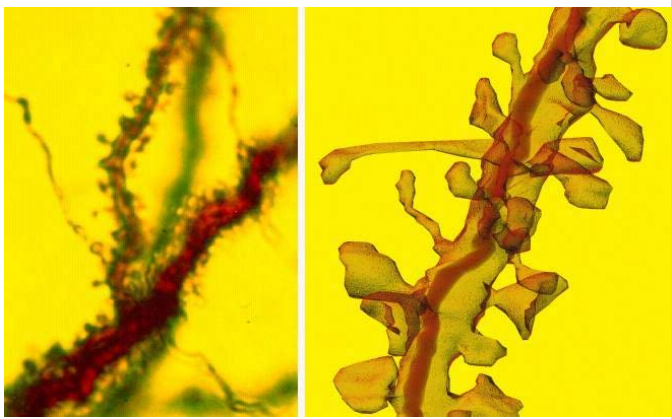

**Figure 13.** Spiny dendrites from hippocampal pyramidal neuron. Left: Light microscope image. Right: Reconstruction from serial electron microscopy.
(From: Synapse Web, Boston University, http://synapses.bu.edu/.)

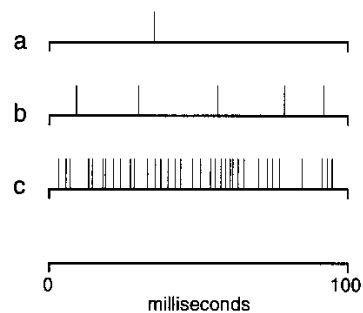

**Figure 15.** The firing of a single neuron. Each short vertical line represents a spike. In (a) the neuron is firing at its background rate. In (b) it is responding to some relevant input by firing at an average rate. In (c) it is responding about as fast as it can.[5]

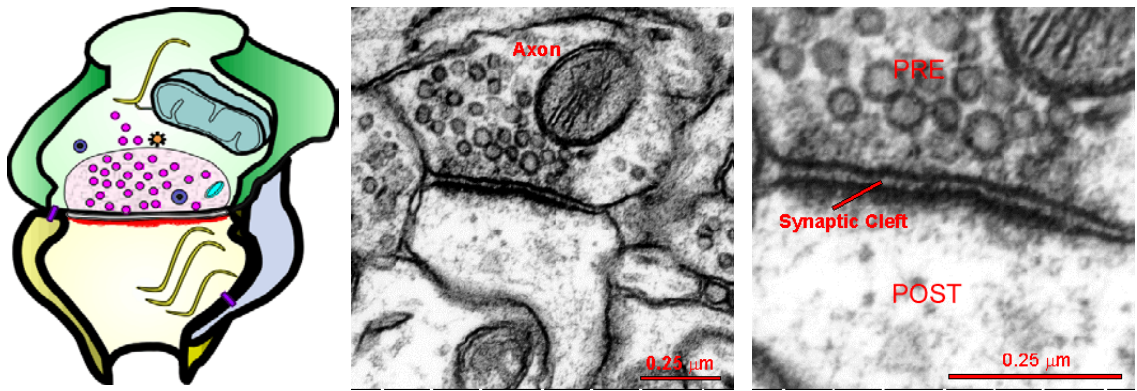

**Figure 14.** (a) Schematic of a synapse. (b) Presynaptic axon and postsynaptic dendritic spine. (c) The synaptic cleft, into which neuro-transmitter is released, is the narrow gap that separates the axonal bouton (PRE) from the postsynaptic cell (POST). (From: Synapse Web, Boston University, http://synapses.bu.edu/)

Broadly speaking, there exist two kinds of neurons: excitatory and inhibitory; see Fig. 16. They have different neuro-transmitters (e.g., the amino acid glutamate in excitatory nerve cells, GABA in inhibitory cells). In the human brain roughly 20% of all neurons are of the inhibitory type. The receiving neuron will make a decision as to whether or not to fire a signal down its axon, based on the collection of signals it receives at its input (dendritic) terminals. The signals coming into a neuron from excitatory cells tend to enhance the possibility of its firing, while the inhibitory signals hamper such chances. The final decision as to whether or not to fire is based on the overall pattern of the inputs; this involves a nonlinear process, which polls the multitude of input signals, performs a complex weighted averaging on them (positive coefficients for excitatory inputs, negative for the inhibitory ones), and if the weighted average exceeds a certain threshold the nerve cell will fire, otherwise it remains in its resting state.[27,28]

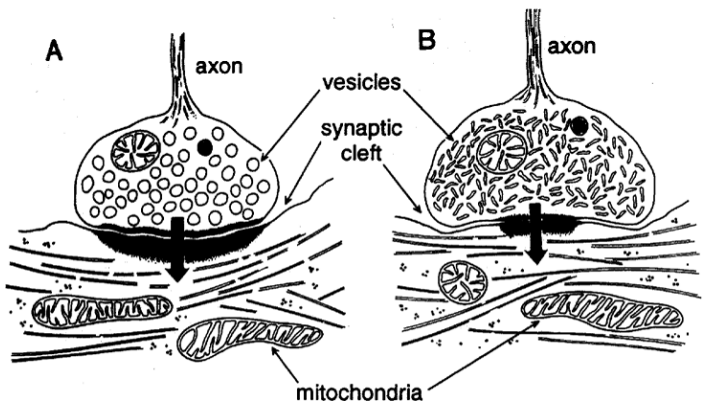

**Figure 16**. The two main types of synapses found in the cortex. A, Type I (excitatory); B, Type 2 (inhibitory). In each figure the axon is at the top, the dendrite at the bottom, with the synaptic cleft in between. The arrows show the direction of the main flow of information. From the (presynaptic) axon to the (postsynaptic) dendrite.[5]

The workings of a neuron in terms of the physical processes of generating an electrical signal, releasing neuro-transmitters, etc., are fairly well understood. As for the mechanism of memory, a model suggested as early as 1945 by the Canadian psychologist Donald Hebb[29] postulates that the synaptic connection between two neurons is strengthened as a result of a successful firing. In other words, if neuron A sends a signal to neuron B, and B decided to fire after polling its inputs, somehow the coincident firings of A and B will affect the strength of the synaptic coupling between the two, making it stronger if A happens to be an excitatory neuron, and reducing the synaptic strength if A happens to be inhibitory. Presumably, the reverse happens if B does not fire in response to A. Human memory, therefore, seems to be formed in the rich pattern of interconnections among billions of neurons in the cortex, by the adjustment of the synaptic strengths.

**The Olfactory System** (Adapted from [28]). The complex interaction among interconnected neurons extending from an external sensory organ to the deepest recesses of the brain is exemplified by the olfactory system. When an animal sniffs an odorant, molecules carrying the scent are captured by a few of the immense number of receptor neurons in the nasal passages; the receptors are somewhat specialized in the kinds of odorants to which they respond. Cells that become excited fire pulses, which propagate through axons to the olfactory bulb (see Fig. 17). The number of activated receptors indicates the intensity of the stimulus, and their location in the nose conveys the nature of the scent. Each scent is expressed by a spatial pattern of receptor activity, which in turn is transmitted to the bulb.

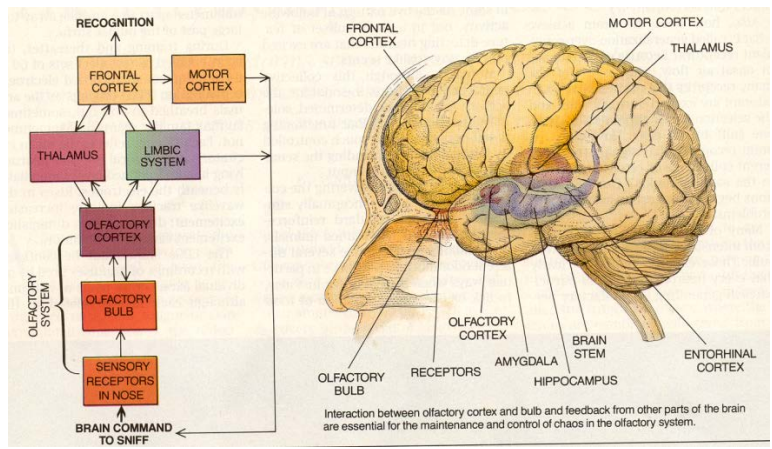

**Figure 17**. Basic flow of olfactory information in the brain.[28]

The bulb analyzes each input pattern and transmits its own message to the olfactory cortex. From there, new signals are sent to many parts of the brain, including the entorhinal cortex, which combines the signals with those from other sensory systems. The result is a perception that is unique to each individual; a perception that cannot be understood solely by examining individual neurons, but depends on the global activity of millions of neurons throughout the cortex.

Cortical neurons continuously receive pulses from thousands of other neurons. Certain incoming pulses generate excitatory waves of electric current in the recipients; others generate inhibitory waves, as shown in Fig. 18. These dendritic currents are fed through the cell body to the "trigger zone" at the start of the axon. There the currents cross the cell membrane into the extracellular space. As they do, the cell calculates the overall strength of the currents by adding excitatory currents and subtracting inhibitory ones. If the sum is above a threshold level, the neuron fires.

By attaching electrodes to an area of the bulbar surface, researchers can collect sets of simultaneously recorded electroencephalogram (EEG) as the animal breathes in and out. Each tracing reflects the mean excitatory state of local pools of neurons lying in a layer beneath the electrodes. Rises in the EEG amplitude indicate increasing excitement; dips represent diminished excitement caused by inhibition. Each EEG is related to the firing pattern of neurons in a neighborhood of the cerebral cortex. The tracings detect essentially the same information that neurons assess when they "decide" whether or not to fire impulses, but an EEG records that information for thousands of cells at once.

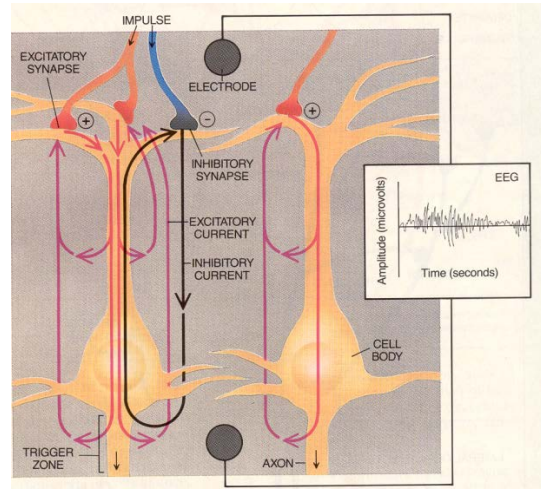

**Figure 18**. EEG waves reflect the mean excitation of pools of neurons. Excitatory inputs at synapses generate electric currents that flow in closed loops within the recipient neuron toward its axon, across the cell membrane into the extracellular space and, in that space, back to the synapse (*red arrows*). Inhibitory inputs generate loops moving in the opposite direction (*black arrows*). In cells the trigger zone adds current strengths (reflected in changes in voltage across the membrane), and it fires impulses if the sum is sufficiently positive. Electrodes on the brain tap those same currents after they leave the cell. The resulting EEGs indicate the excitation of whole groups of cells, not individuals, because the extracellular avenues from which the EEGs arise carry currents contributed by thousands of cells.[28]

As shown in Fig. 18, the mechanism producing each EEG tracing sums the currents initiated at the dendrites, but it taps the currents after they leave the cell. The tracings reflect the excitatory state of groups of neurons rather than of individual ones, because the extracellular space is traversed by currents from thousands of cells.

In living individuals, EEGs always oscillate to some extent, but the oscillations are usually quite irregular. When an animal inhales a familiar scent, a burst can be seen in each EEG tracing. All the waves from the array of electrodes suddenly become more regular for a few cycles – until the animal exhales (see Fig. 19). The waves often have a higher amplitude and frequency than they do at other times. (The bursts are often called 40 Hz waves, although the frequency can actually range from 20 to 90 Hz.)

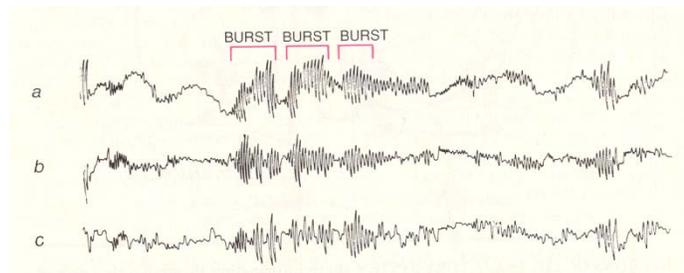

**Figure 19**. Simultaneous recordings from the olfactory bulb (a) and front (b) and rear (c) parts of a cat's olfactory cortex show low-frequency waves interrupted by bursts – high-amplitude, high-frequency oscillations that are generated when odors are perceived. The average amplitude of a burst is ~ 100 μV. Each lasts a fraction of a second, for the interval between inhalation and exhalation.[28]

Curiously, it is not the shape of the carrier wave that reveals the identity of an odor. Indeed, the wave changes every time an animal inhales, even when the same odorant is repeatedly sniffed. The identity of an odorant is discernible only in the bulbwide spatial pattern of the carrier-wave amplitude. It is believed that something called "the nerve cell assembly" is both a crucial repository of past associations and an essential participant in the formation of the collective bulbar burst. The hypothetical assembly consists of neurons that have simultaneously been excited by other neurons during learning.

When animals are trained to discriminate among olfactory stimuli, certain synapses that connect neurons within the bulb and within the olfactory cortex become selectively strengthened during training. That is, the sensitivity of the postsynaptic cells to excitatory input is increased at the synapse, so that an input generates a greater dendritic current than it would have generated in the absence of special training.

The strengthening occurs not in the synapse between an input axon (e.g., a nasal receptor) and the neuron it excites (e.g., a bulbar neuron), but in the connection between neurons that are simultaneously excited during learning. Neurons in the bulb and in the olfactory cortex are connected to many others in those regions. Such strengthening is predicted by the Hebb rule, which holds that synapses between neurons that fire together become stronger, as long as such synchronous firing is rewarded. (Strengthening involves modulator chemicals released by the brain stem during reinforcement.)

Thus a nerve cell assembly, consisting of neurons joined by Hebbian synapses, forms for a particular scent as the individual learns to identify that odorant. Subsequently, when any subset of neurons in the assembly receives the familiar input, the entire assembly is stimulated, as excitatory signals race across the favored Hebbian synapses. The assembly, in turn, directs the rest of the bulb into a distinct pattern of activity.

If the odorant is familiar and the bulb has been primed by arousal, the information spreads like a flash fire through the nerve cell assembly. First, excitatory input to one part of the assembly during a sniff excites the other parts, via the Hebbian synapses. Then those parts reexcite the first, increasing the gain, and so forth, so that the input rapidly ignites an explosion of collective activity throughout the assembly. The activity of the assembly, in turn, spreads to the entire bulb, igniting a full-blown burst. The bulb then sends a "consensus statement" simultaneously along parallel axons to the olfactory cortex (see Fig. 20).

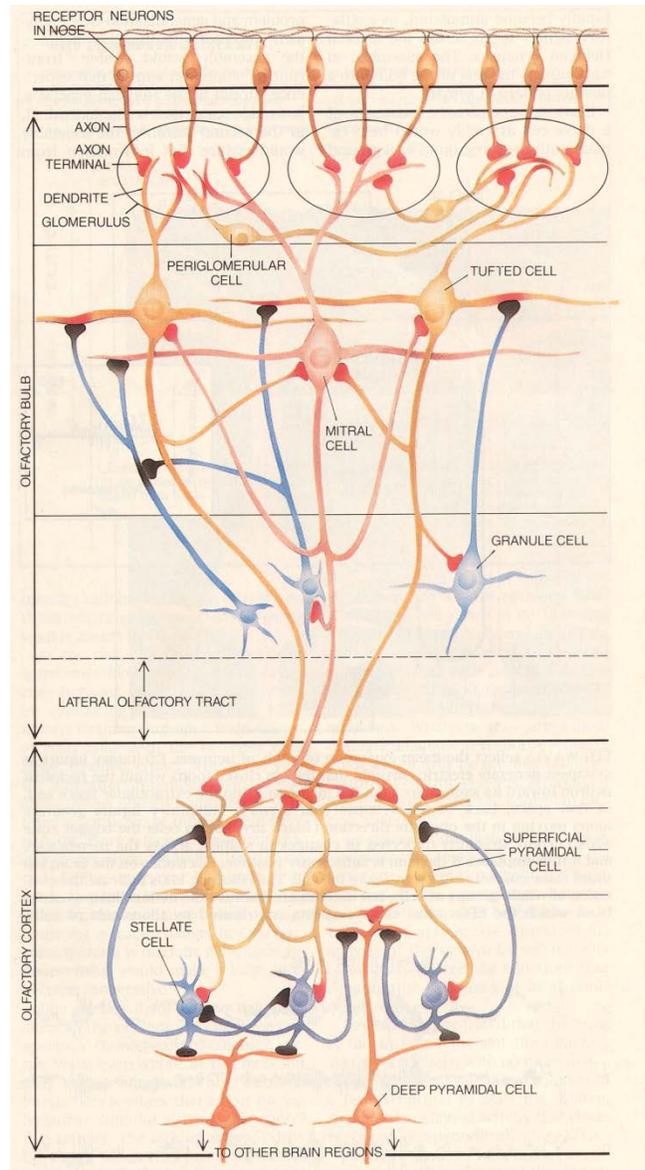

**Figure 20**. Neurons of the olfactory system share information through a rich web of synapses, junctions where signals flow from neuron to neuron. Usually signals pass from projections called axons to projections called dendrites, but sometimes they pass from dendrite to dendrite or axon to axon. The widespread sharing leads to collective activity. In this highly schematic diagram, red shading signifies that a neuron is exciting another cell, black shading that a neuron is inhibiting another.[28]

How does that cortical area distinguish the consensus statement from the background of other stimuli? The answer has to do with the wiring that joins the bulb to the cortex. The bulb generates trains of impulses that run simultaneously along the parallel axons leading from the bulb to the cortex. Each axon branches extensively and transmits pulses to many thousands of neurons across the olfactory cortex, and each cortical target cell receives input from thousands of bulbar cells. The carrier activity of the incoming lines, synchronized by cooperation, probably stands out for the simple reason that such signals add together. Nonsynchronous inputs, which are not at the carrier frequency and phase, effectively cancel one another.

Thus, every recipient neuron in the olfactory cortex picks up a share of the cooperative bulbar signal and transmits the summed signals to thousands of its neighbors simultaneously. In response, the cortical neurons, which have formed nerve-cell assemblies, promptly generate their own collective burst, albeit one having a spatial pattern that differs from those in the bulb. Just as a burst in the bulb guarantees the delivery of a coherent message to the cortex, the global burst in the cortex enables outgoing messages from that region to stand above the din when they reach other regions of the brain.

**The Hierarchy of Cell Assemblies** (adapted from [26]). Donald Hebb first entertained the cell assembly theory in the mid-1940s after reading Marius von Senden's *Space and Sight*, originally published in 1932. In this work von Senden assembled existing records on 65 patients from the eleventh century up to the year 1912 who had been born blind due to cataracts. At ages varying from 3 to 46, the cataracts were removed, and a variety of reporters observed them as they went about handling the sudden and often maddeningly novel influx of light.

One of the few uniformities in these cases was that the process of learning to see "is an enterprise fraught with innumerable difficulties, and that the common idea that the patient must necessarily be delighted with the gifts of light and color bequeathed to him by the operation is wholly remote from the facts." Not every patient rejoiced upon being forced to make sense of light that was all but incomprehensible, and many found the effort of learning to see so difficult that they simply gave up. The doctor of a 21 year-old woman reported:

> Even before the operation she was dull and spiritless. Whenever I visited her afterwards, the bandages having been long since discarded, I found her with her eyes closed, though not forced to do so by any aversion to the light. Laborious persuasion was needed before she would even come to look at the things immediately about her, and at last to become acquainted with them. Some years ago, her father wrote that she carefully shuts her eyes whenever she wishes to go about the house, especially when she comes to a staircase, and that she is never happier or more at ease than when, by closing her eyelids, she relapses into her former state of total blindness.

If seeing were a simple reflex, the restoration of vision should have completed a circuit, allowing a patient to view the world as lucidly as if he or she had done so all along. Instead von Senden's accounts suggest that seeing was not at all automatic. The friend of one 15 year-old patient, in an attempt to encourage him to learn to see, tried the following experiment:

> One day, during the vine-harvest, she picked a bunch of grapes and showed it to him from a distance. "What is that?" she asked. "It is dark and shiny," he replied. "Anything else?" "It isn't smooth, it has bumps and hollows." "Can one eat it?" she asked. "I don't know." As soon as he touched the bunch, he cried: "But they're grapes!"

The reports indicated that patients who had been completely dependent on tactile impressions before the operation, had an awareness of space that was totally different from a normal visual awareness. At first, von Senden discovered, "the patient feels visual impressions to be something alien, intruding on his mind without action on his own part." Later, he reported, "the stimuli impinging on the visual organ from an objective shape merely occasion the act of perception as such, but do not determine its outcome." A 9 year-old boy, for example, spent days trying to learn how to tell a sphere from a cube. From his record:

> He gradually became more correct in his perception, but it was only after several days that he could tell by his eyes alone, which was the sphere and which the cube. When asked, he always, before answering, wished to take both into his hands. Even when this was allowed, when immediately afterwards the objects were placed before the eyes, he was not certain of the figure.

That such observations are not artifacts of the surgery or uniquely human was established through observations on a pair of chimpanzees that had been reared in the dark by a colleague of Hebb. After being brought out into the light, these animals showed no emotional reactions to their new experiences. They seemed unaware of the stimulation of light and did not try to explore visual objects by touch. Hebb concluded that the chimps showed no visual response because they had not formed the neural assemblies that are necessary for visual perception.

"How," Hebb asked, "can it be possible for a man to have an IQ of 160 or higher, after a prefrontal lobe has been removed?" If a particular memory is stored in a certain neuron, its continued existence depends on the life of the cell. Cell death, in this model, means total loss of the memory. Assembly storage, on the other hand, is robust because physical damage – leading to the death of a few cells – will not necessarily destroy specific assemblies but instead degrade many by roughly equal degrees. This loss could then be restored through recruitment of additional cells into the assembly.

If one accepts the concept of cell assemblies, it is interesting to ask how many might be able to form. This is a difficult question, but Charles Legéndy has obtained some results from a simple model. He assumes that the brain is already organized into subassemblies and discusses their organization into larger assemblies. The assembly and one of its subassemblies variously represent "a setting and a person who is part of it, a word and one of its letters, an object and one of its details."

In Legéndy's model, interconnections are assumed to be evenly distributed over the neocortex to avoid the complications of spatial organization. His subassemblies are formed through weak contacts, and assemblies emerge from subassemblies through the development of latent into strong contacts between neurons. From statistical arguments, Legéndy estimates the maximum number of assemblies in a brain to be about $(N/ny)^2$, where $N$ is the number of neurons in the brain, $n$ is the number of neurons in a subassembly, and $y$ is the number of subassemblies in an assembly. Taking $N = 10^{10}$, $n = 10^4$, and $y = 30$, he finds ~$10^9$ assemblies, as a conservative estimate for "the number of elementary things the brain can know." The number of neurons in the neocortex may be an order of magnitude greater than $10^{10}$, and the real neuron is much more complex than the simple model assumed by Legéndy, but $10^9$ is approximately the number of seconds in 30 years. Thus a model of the brain based on simple neurons provides sufficient storage for the complex memories of a normal lifetime.[30]

**Cortical Stimulation** (adapted from [26]). The Canadian neurosurgeon W. Penfield and his colleagues induced hallucinations in several patients through electrical stimulation of the neocortex.[31] The purpose of such stimulation was to locate the origin of epileptic activity in order to remove the offending portion of the cortex. Gentle electrical stimulation was applied to the temporal lobes of 520 patients, of whom 40 reported experiential responses. Stimulating currents between 50 and 500 µA were used in pulses of 2 to 5 ms at frequencies of 40 to 100 Hz.

M.M. was a typical case. A woman of 26, she had her first epileptic seizure at the age of 5. When she was in college, the pattern included visual and auditory hallucinations, coming in flashes that she felt she had experienced before. One of these

> had to do with her cousin's house or the trip there – a trip she had not made for ten to fifteen years but used to make often as a child. She is in a motor car which had stopped before a railway crossing. The details are vivid. She can see the swinging light at the crossing. The train is going by – it is pulled by a locomotive passing from the left to right and she sees coal smoke coming out of the engine and flowing back over the train. On her right there is a big chemical plant and she remembers smelling the odor of the chemical plant.

During the operation, her skull was opened and her right temporal region explored to locate the epileptic region. Figure 21 shows the exposed lobe with numbered tickets that mark the sites at which positive responses were evoked. Penfield termed the responses *experiential* because the patient felt that she was reliving the experience – not merely remembering it – even as she remained aware that she was lying in the operating room and talking to the doctor. The following experiential responses were recorded upon electrical stimulation at the numbered locations:

11. She said, "I heard something familiar, I do not know what it was."

11. Repeated without warning. "Yes, sir, I think I heard a mother calling her little boy somewhere. It seemed to be something that happened years ago." When asked if she knew who it was she said, "Somebody in the neighborhood where I live." When asked she said it seemed as though she was somewhere close enough to hear.

11. Repeated 18 minutes later. "Yes, I heard the same familiar sounds, it seems to be a woman calling. The same lady. That was not in the neighborhood. It seemed to be at the lumber yard."

13. "Yes, I heard voices down along the river somewhere – a man's voice and a woman's voice calling." When asked how she could tell it was down along the river, she said, "I think I saw the river." When asked what river, she said, "I do not know, it seems to be one I visited as a child."

13. Repeated without warning. "Yes, I hear voices, it is late at night, around the carnival somewhere – some sort of a travelling circus. When asked what she saw, she said, "I just saw lots of big wagons that they use to haul animals in."

12. Stimulation without warning. She said, "I seemed to hear little voices then. The voices of people calling from building to building somewhere. I do not know where it is but it seems very familiar to me. I cannot see the buildings now, but they seemed to be run-down buildings."

14. "I heard voices. My whole body seemed to be moving back and forth, particularly my head."

14. Repeated. "I heard voices."

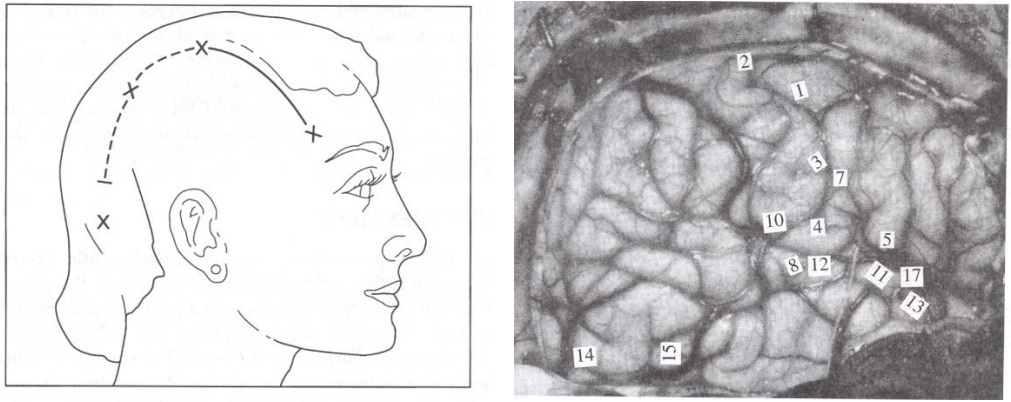

**Figure 21**. The right temporal lobe of M.M. with numbers that indicate points of positive stimulation.[31]

The forty cases surveyed by Penfield and his colleagues show that this particular example of experiential response is not at all unique. They conclude that, in the human brain, there is a remarkable record of the stream of each individual's consciousness, and that stimulation of certain areas of the neocortex – lying on the temporal lobe between the auditory and visual sensory areas – causes previous experience to return to the mind of a conscious person.[26]

**Information in Human Societies**. The information used by humans is of several kinds. We write down recipes and instruction sets for making things such as high-rise buildings, apple pies, DVD-players, clothing items, chemicals, movies, etc. Musicians compose new music, playwrights write original plays, painters create new paintings, mathematicians discover new theorems, scientists uncover facts about Nature, and legislators create new laws. These ideas and expressions (rendered in various media) need to be stored and protected for use by other members of the society, or for transmission to future generations. Daily events, stock prices, sports news, and weather information are likewise useful and often worthy of short-term or long-term storage. A culture is shaped by the collection of all such information learned, created, applied, imitated, modified, improved upon, and transmitted within a population and from one generation to the next.

Within the human brain, a piece of information is more than just a rigid set of connections among certain neurons. The sounds, images, scents, events, ideas and intuitions residing within the confines of a given skull are interconnected. In some instances they help to reinforce one another, in others they give rise to logical inconsistencies and absurdities that the conscious mind attempts to resolve. Once resolution occurs some of the memories gain prominence at the expense of others. Mental activity of this kind often results in the creation of new concepts and ideas, which themselves explore the content of the brain and attempt to form new connections. This is a dynamic, self-propelled environment, which is constantly gaining knowledge, combining and restructuring the existing memories, discounting or forgetting some previously-held beliefs, and initiating actions that result in new explorations, further acquisition of information, and so forth.

Compare the above situation with the passive nature of the information stored in a book or on a hard drive. By and large the various pieces of information are ignorant of each other, their interactions are either non-existent, or mediated by the mind of a user or by the CPU of a computer. Our data storage technology is divorced from the processing technology that operates on the stored information. Our stored information cannot initiate a search for other pieces of related data, cannot resolve its internal inconsistencies, cannot form alliances with other pieces of data sitting on the same medium that might bolster its effectiveness, cannot create new pieces of information or improve its own content. Such are the shortcomings of the storage technology at the end of the 20$^{th}$ century. Although the search for denser, faster, and cheaper memories is a worthy goal for now and for many years to come, it is far more challenging (and proportionately more rewarding) to search for alternative modes of storage that would overcome the passive nature of the present-day technology in the broad sense just described.

To the lively and evolving environment within a human brain, now add the interactions among billions of similar brains on the planet. An idea formed in one head moves to several other heads, undergoes modifications, forms new alliances, then propagates further to others. Spoken language was perhaps the first extensive means of inter-brain communication, followed by the written word. Then came the printing press in the 1440's and revolutionized human lives and cultures. Electronic communication (radio, TV, phone, fax, copy machines) has had a profound effect on our lives over the past century, but perhaps all these revolutions will be dwarfed by the promise of the internet and what it will do for the inter-brain connectivity.

The internet is not just allowing us human beings to access all the knowledge of the past and the present, and to explore each other's minds. The internet is also allowing computers and other electronic means of information processing and storage to communicate among themselves. At the moment this communication is mediated by humans: All inter-computer communication is under the control of software written by engineers and enthusiasts. But imagine the day when all the computers of the world, connected via the internet, can exchange their information and form data structures above and beyond those imposed upon them by their human masters. Imagine the day when control programs, recipes, musical scores, movie scripts, mathematical theorems, etc., can mutate, mate, replicate, propagate through the internet, and survive or perish in the competition with their kin or with other denizens of the net. In their constant struggle for survival, only those memes[32] that are best adapted to their natural environment will emerge and flourish. (Meme is the unit of cultural inheritance; see the glossary for further details.)

**Does Nature Follow any Principles of Organization?** Darwin's principles of evolutionary biology[33] are quite simple, yet they have provided a powerful framework for the study of life processes and their historical development. First there is the inheritance of biological characteristics by the offspring from parent(s). Second, genetic mutation can change some of these transmitted characteristics in apparently random ways. Third, there is natural selection based on the survival of the fittest individuals.[2] The genetic code, while transmitted from one generation to the next, remains largely intact, although paternal and maternal chromosomes are shuffled rather extensively, and a few genes here and there are modified in the course of transmission to the next generation.[10] If these modified genes endow the offspring with useful qualities – qualities that help survival and reproduction – then they remain in the gene pool and spread throughout the population, otherwise the individual will die or will fail to reproduce, causing the disappearance of those (harmful) mutated genes.

The Darwinian principles do not explain the hierarchical organization and the "sense of purpose" that seem to guide the processes of evolution. Are there additional rules that Nature follows in directing the evolution of species? The complex pattern of interconnection among the brain cells, to pick but one example, is not likely to have resulted from random trial and error;[34,35] there are just too many possibilities for interconnection and too little time (~ 600 million years since the appearance of multicellular organisms) for these connections to have been tried in all their varieties before being adopted. What rules does Nature follow in allowing collective mutations, for example? To what extent and under what circumstances is repetition (e.g., gene duplication) used by Nature? What kind of mutations in duplicated genes are most beneficial? We do not know the answers to these questions; in fact, we probably have not even begun to ask the right questions. Perhaps the time is now ripe to begin to focus on such issues, as the underlying structure and functioning of the cells and the electrochemical basis of intercellular communication have already been worked out in great detail during the past century. Our understanding of the molecular basis of life has reached a point that one can now ask questions about higher-level principles of organization with regard to the architecture of complex biological systems. It is in asking and attempting to answer such questions that humanity faces one of its greatest challenges today.

**Observations Concerning Higher-level Principles of Organization**. Here is one set of principles that the Nature seems to have consistently used throughout the biosphere.

(i) There exist simple functional or operational units (hereinafter referred to as action units), that appear over and over again within an organism, or even cross the boundaries of individual organisms to appear across large groups of individuals and species. Examples include:

- Expansion or contraction of a muscle cell in response to a command from a motor neuron.[25]
- Splitting into two parts and forming new functional units similar in constitution to the parent (i.e., cell division, DNA duplication).
- Random shuffling (e.g., chromosome recombination during meiosis to endow one's offspring with a mixture of traits from one's own parents;[10] gene shuffling in the immune cells for producing a large number of receptor molecules to match a multitude of foreign antigens[6]).

These "fundamental" units of biological action are activated in response to a set of input signals, and result in specific outputs or actions. The input signals may vary over a limited range without affecting the output, or they may affect the output quantitatively but not qualitatively. In other words, the basic action units have some degree of flexibility, tolerating variations in the level of the electrical/chemical input signal(s), relative timing of these signals, etc.

(ii) The various action units communicate and interact with each other. One action can trigger another, and in turn be affected by the result of the action it triggers. In other words, inputs to each unit could be the outputs from other units. Biological means of communication include:

- Cell-to-cell communication by chemical means, e.g., via neurotransmitters and hormones.
- Cell-to-cell communication by electrical signaling, e.g., from neuron to neuron, from neuron to muscle cell and vice versa, and from neuron to glandular cell.
- Migration of individual cells through the blood circulation system, through the lymphatic system, or by crawling over other cells.
- Directional growth and branching of individual cells to establish contact with distant cells.[34]

The interaction between two units may occur simultaneously, or it may occur sequentially in the sense that the response of one unit will follow the action of another. Some actions are triggered by external causes (e.g., variations in temperature or pressure, influx of light, arrival of chemicals from the outside world), while others are initiated by internal signals produced by some other action unit(s).

(iii) Individual action units have the potential to switch from one mode of operation to another, depending on their internal state and the nature of the incoming signals. For example,

- Motor neurons can grow additional branches from their axons and make new contacts to the muscle cells if some of their neighbors die (e.g., in polio survivors), or if they are stimulated to do so by exercise.
- The fertilized eggs of many reptiles (e.g., turtles, various lizards) grow into males or females depending on their ambient temperature.
- Various stem cells can grow into different types of cell, depending on their mode of stimulation and chemical environment. The stem cells of the immune system, for instance, originate in the bone marrow and, after several divisions, are transformed into erythrocytes (red blood cells) or lymphocytes (white blood cells, such as T-cells, B-cells, phagocytes, etc.) depending on whether they mature in thymus or bone marrow, and also depending on the chemical signals they receive prior to differentiation and maturation.[6]

This degree of flexibility is supported by the fact that each cell within the organism carries the entire genetic code of that organism, making it possible to switch from one instruction set to another when prompted by the circumstances in which the cell might find itself.

(iv) The strength of interaction between two action units could be modified depending on the outcome of their direct interaction (e.g., the Hebbian model for adjustment of the synaptic strength), or as a result of internal modifications to individual units (i.e., an increase or decrease in the amount of available neurotransmitter molecules), or as a result of a change in the environment. Moreover, two disjoint units could develop interactions (for example, by growth and connection of new axonal branches), or two interacting units may lose their connections over time.

We do not yet know how Nature decides to establish connections among various action units. Does the Nature follow any specific rules in deciding whether two units should or should not interact? Since the underlying network of interactions (or the propensity of individual units to grow towards others, make contact with them, regulate the strength of their interactions, etc.) are controlled by the genes, one should ask whether there are any features of chromosomal recombination during meiosis, gene duplication, and gene transfer, etc., that could be responsible for promotion of such behavior among the individual action units. Once this underlying network, and the guiding principles for its growth and connectivity are established, each individual organism responds to its unique environment (and its unique upbringing) by developing special connections among its constituent action units. The individual thus forms its own memories, talents, personality, etc.[30] None of these acquired features, of course, are heritable and, therefore, they do not induce changes in the genome of the species.[10] But they help the individual to prosper and to survive long enough to transfer to its offspring the same genes that endowed him/her with those advantageous traits.[33]

For the technologist seeking to imitate Nature, the task ahead may thus be summarized as follows: Identify the genetic mutation rules and patterns that enable the formation, growth, and development of individual action units, which tend to seek and establish "beneficial" interactions with other (similar or dissimilar) units. Needless to say, deciding what constitutes a beneficial interaction may, in itself, be a formidable task.

**Conclusion**. Nature has evolved certain biological systems that are capable of processing and storage of information far more efficiently than any man-made device produced to date can accomplish. Progress in molecular biology and fundamental understanding of embryology, immune system function, the genetic basis of inheritance, etc., have created an impressive knowledge-base pertaining to the functioning of individual cells and many of their interactions.[36-39] What seems to be lacking is an understanding of the principles of development of interaction among various units, and the organizational principles that allow useful interactions to develop, while providing flexibility for interconnections to modify themselves according to the environmental conditions.[40] These underlying principles are believed to be simple and elegant. Just as the Darwinian principles of evolutionary biology have provided a framework and a guiding light for biologists, the heretofore undiscovered principles of organization among the fundamental biological action units will enable information technologists to create new methods of information storage/processing that promise to be far more powerful than any method that is predicated on existing technologies.

# Glossary

**Allele**. One of the variant forms of a gene at a particular locus, or location, on a chromosome. Different alleles produce variation in inherited characteristics such as hair color or blood type. In an individual, one form of the allele (the dominant one) may be expressed more frequently than another form (the recessive one).

**Axon**. The output cable of a neuron. A neuron usually has only one axon, although it often branches extensively.

**Basket cell**. In the cerebral cortex, a type of inhibitory nerve cell, often with a rather long axon, that extends to make multiple contacts on or near the cell bodies of other neurons.

**Cerebral cortex**. Often called simply "the cortex." A pair of large folded sheets of nervous tissue, one on either side of the top of the head. It is sometimes divided into three main regions, called the neocortex (the largest part in primates), the paleocortex, and the archicortex.[5]

**Corpus callosum**. A very large tract of nerve fibers (axons), connecting the two halves of the cerebral cortex.[5]

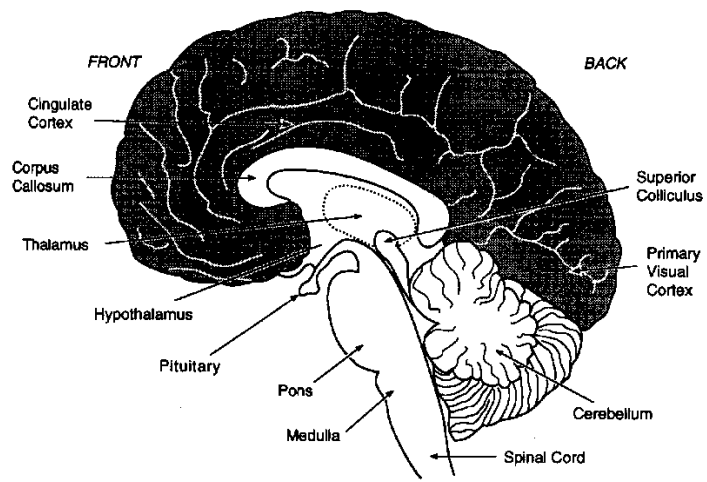

Major parts of the human brain, viewed from the inside. The cortex is shaded gray.[5]

**Chromosome**. One of the threadlike packages of genes and other DNA in the nucleus of a cell. Different organisms have different numbers of chromosomes. Humans have 23 pairs of chromosomes, 46 in all: 44 autosomes and two sex chromosomes (XX in females, XY in males). Each parent contributes one chromosome to each pair, so children get half of their chromosomes from their mothers and half from their fathers. Human beings have ~ 3.2 billion base pairs and ~34,000 genes in their 23 chromosome pairs. These genes are estimated to give rise to between 500,000 and 1,000,000 different proteins. The fruit-fly genome has 185 million base pairs and 13,601 genes. The genome of the roundworm has 97 million base pairs and 19,099 genes.

**Dendrite**. A treelike part of a nerve cell, dendrites receive signals from other nerve cells.

**DNA (deoxyribonucleic acid)**. Built from nucleic acid bases, sugars, and phosphates, the double-stranded molecule is twisted into a helix. Each spiraling strand, comprised of a sugar-phosphate backbone and attached bases, is connected to a complementary strand by non-covalent hydrogen bonding between paired bases. The bases are adenine (A), thymine (T), cytosine (C) and guanine (G). A and T are connected by two, and G and C by three hydrogen bonds. The right-handed double helix has ~10 nucleotide pairs per helical turn. The double-helix structure of DNA was discovered in 1953 by James Watson and Francis Crick.

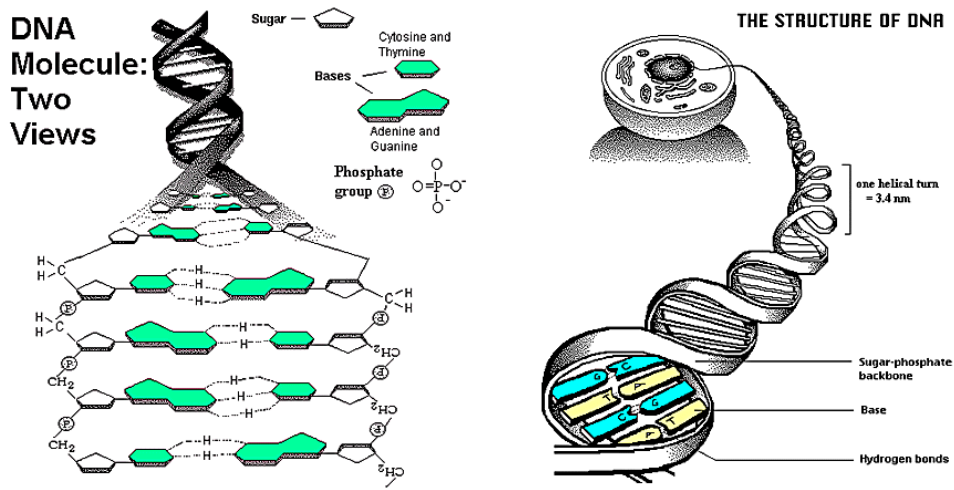

(Reproduced with permission from: http://www.accessexcellence.org/ )

**Enzyme**. A protein, made from amino acids, acting as a catalyst in the intracellular chemical processes. In almost all cases enzymes are very large protein molecules, although some have smaller organic molecules attached to them. Enzymes help break down external molecules that are useful to the cells, and increase the chemical rates of interaction between various organic molecules. For example, the attachment of amino acids to transfer RNA (tRNA) is mediated by certain enzymes. (A catalyst accelerates a chemical reaction, but is unchanged at the end of it.)

**Eukaryote**. Eukaryotes include protozoa, green algae, yeasts, fungi, and all higher plants and animals. Eukaryotic cells have a nucleus and nuclear membrane. Their chromosomal DNA is permanently associated with a variety of proteins. The cells contain mitochondria and, in algae and higher plants, chloroplasts.[9,10]

**GABA**. A small chemical whose proper name is gamma-amino-butyric acid. It is the major inhibitory neurotransmitter in the forebrain.[5]

**Glutamate**. A small organic chemical, the major excitatory neurotransmitter in the forebrain.[5]

**Hebb's rule**. Named after Canadian psychologist Donald O. Hebb. A type of alteration to the strength of a synapse that depends on both the presynaptic activity coming into the synapse and the activity (of some sort) of the receiving neuron on the postsynaptic side. It is important because the alteration to the synapse requires the association in time of two distinct forms of neural activity.[5]

**Lipid**. A general descriptive term for certain organic molecules that have one water-loving end and one fat-loving end. A double layer of lipid makes up the lipid bilayer that forms the basis of most biological membranes, such as that which surrounds each cell.[5]

**Meiosis**. The process of cell division that results in the production of sperm or egg cells (gametes). During meiosis (lessening) each cell splits twice, but the chromosomes split only once, resulting in sperm or egg cells that contain only one chromosome from each pair of chromosomes present in the original cell. The rules of Mendelian inheritance, however, cannot explain the behavior of linked genes (i.e., genes at different loci on the same chromosome). The relevant facts are illustrated in the figure below (only a single pair of homologous chromosomes is shown). The important events are as follows:

(a) Each chromosome is replicated, so that it consists of two identical threads.
(b) Pairs of homologous chromosomes come to lie side by side, forming bivalents. Each bivalent then consists of four similar threads.
(c) The two members of a pair repel one another, but are held together at a few points, called chiasmata; there are two such chiasmata in the bivalent depicted in the figure.
(d) There are two successive divisions of the nucleus, without further chromosome replication, giving rise to four nuclei (the gametic nuclei), each containing a single set of chromosomes.

In the figure, the chromosome derived from the father (paternal) is shown cross-hatched, to distinguish it from the maternal chromosome. In the four-strand and later stages, sections of chromosome which are copies of the original paternal chromosome are likewise shown cross-hatched, although it is not normally possible to distinguish maternal and paternal chromosomes under the microscope. The genetic consequence of all this is that if two genes are inherited from the same parent, they tend to be transmitted together, but if a recombination takes place between them, one is transmitted without the other.[10]

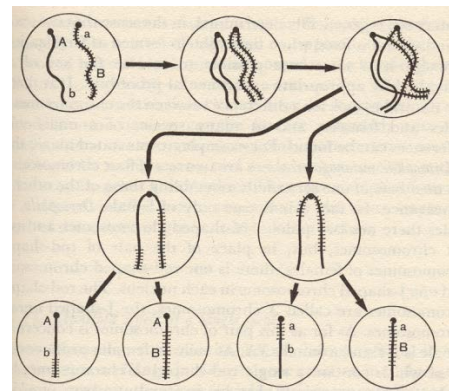

Recombination of a pair of 'homologous' chromosomes during meiosis.[10]

**Meme**. A word coined by Richard Dawkins in his best-selling book, *The Selfish Gene*.[2] Meme is the unit of cultural inheritance; it copies itself from brain to brain via any available means of copying. Memes can be good ideas, good tunes, memorable jokes, religious beliefs, popular poems, brilliant insights in mathematics, clever techniques for tying knots, etc. Those memes that spread do so because they are good at spreading; they don't have to be good or bad in any practical or utilitarian sense. Memes spread in a population because of the biologically valuable tendency of individuals to imitate. Memes not only leap from mind to mind by imitation in culture, but also they thrive, multiply, and compete within our minds. Whereas genes build the hardware, memes are the software; the coevolution of genes and memes is what may have driven the inflation of the human brain. (Adapted from R. Dawkins, *Unweaving The Rainbow*, Mariner Books, 1998.)

**Mitosis**. The process of cell division. In mitosis both pairs of chromosomes are duplicated, and each daughter cell receives a complete complement of chromosomes from its progenitor.

**Neuron**. The structural unit of the nervous system, contains a nucleus and cell body from which has grown a tree of many branches and twigs spreading in various directions.[25] A neuron typically has many dendrites and one axon. The dendrites branch and terminate in the vicinity of the cell body. In contrast, axons can extend to distant targets, more than a meter away in some cases. Dendrites are rarely more than a millimeter long and often much shorter. The drawing on the right (after Ramón y Cajal, 1952) shows a purkinje cell of the human cerebellum, which receives about 80,000 synaptic inputs.[26]

**Prokaryote**. A cellular organism (as a bacterium or a blue-green alga) that does not have a distinct nucleus or nuclear membrane. The DNA is arranged in a single ring-shaped chromosome with few associated proteins. There are no separate chloroplasts or mitochondria. Prokaryotes are the most primitive (and the most successful) form of life, dating back about 3.2 billion years.[9]

**Protein**. A molecule formed as a linear sequence of amino acids. Although there are many amino acids in nature, only 20 of them are used as the building blocks of proteins. These amino acids are the same in all living organisms. The basic end of one amino

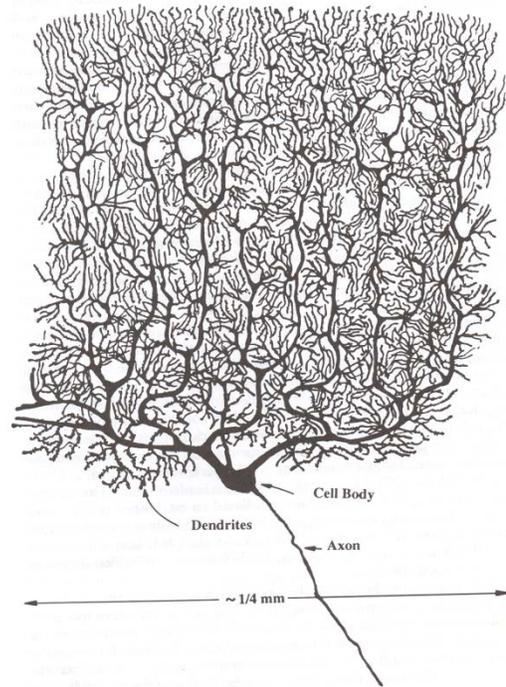

A purkinje cell of the human cerebellum.[26]

acid typically combines with the acidic end of another to form a lasting connection, and in the process releasing a molecule of water. Unlike DNA molecules, all proteins are single stranded. They usually fold on themselves under the influence of electrostatic forces to assume various geometrical shapes. Examples of amino acids used as building blocks of proteins are leucine, valine, proline, tryptophan, methionine, arginine, and phenylalanine.

Hair and skin are made of a protein called keratin, which is also the crucial ingredient in biological materials like feathers and fingernails, hooves and horns. Another protein, collagen, is the stuff of connective tissues such as tendons, bones, and cartilage. Fibroin is the silk of cocoons and spider webs. Sclerolin forms the external skeletons of insects such as the firefly. Such proteins are *structural,* lending shape and form to life. Still other proteins are more *functional,* taking crucial roles in other processes within cells. Insulin is a well-known protein that regulates the metabolism of glucose, a natural sugar. Rhodopsin, in the retina of the eye, converts incoming light to ionic signals in the optic nerve. Myosin and actin provide the forces in vertebrate muscle cells. Dynein is an energy-inducing component of the cytoskeleton. Hemoglobin carries oxygen in the blood, and myoglobin (a close cousin) carries oxygen in muscle. Myoglobin has the chemical formula $C_{738}H_{1166}FeN_{203}O_{208}S_2$; the iron atom forms part of an active site, the heme group, that binds oxygen deep inside the molecule. Almost all enzymes – catalysts that speed biochemical reactions but are not consumed by them – are functional proteins. So are the transmembrane sodium-, potassium-, and calcium-channels in nerve membranes, which support the basic electrochemical activity of the brain.[26]

**Protozoan**. Any of a phylum or subkingdom (Protozoa) of minute protoplasmic acellular or unicellular animals which have varied morphology and physiology and often complex life cycles which are represented in almost every kind of habitat, and some of which are serious parasites of man and domestic animals (Webster's new dictionary). Protozoa are single-celled eukaryotes.

**RNA (Ribonucleic acid)**. A chemical similar to a single strand of DNA. In RNA, the letter U, which stands for uracil, is substituted for T in the genetic code. Messenger RNA delivers DNA's genetic message to the cytoplasm of a cell where proteins are made. Three-letter sequences of messenger RNA, known as codons, code for specific amino acids. For example, UUU and UUC both code for phenylalanine, AUG codes for methionine, and UGG codes for tryptophan. Some amino acids have only one codon, others have more than one (for example, leucine has six). Three codons (UAA, UAG, and UGA) operate as "stops," marking the end of a gene.

For each kind of coded triplet on mRNA, there exists a special transfer RNA molecule (tRNA), which will pair with just that triplet, and also a special activating enzyme, which will attach one kind of amino acid, and only one kind, to that transfer molecule. It is the function of the transfer molecules and activating enzymes to line up the amino acids opposite their appropriate triplets in the messenger molecule. The amino acids are then joined up end to end to form the completed protein. The triplet of bases on a transfer molecule which pairs with a codon on mRNA is called an anticodon.[10]

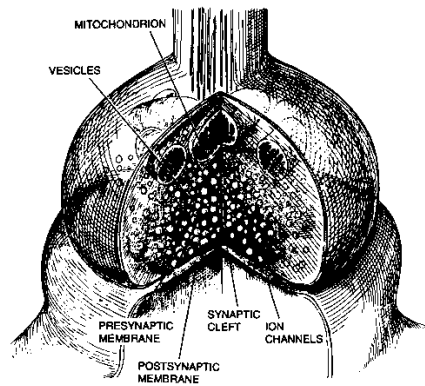

**Synapse**. The connection between one nerve cell and another. Most of these have a minute gap (between the terminal of the incoming axon and the recipient neuron) across which neurotransmitter molecules can diffuse. In some parts of the brain the dendrites of one cell can form a synapse on the dendrites of another, but such synapses are rare or absent in the cerebral cortex.[5]